\documentclass[preprint,12pt]{elsarticle}

\usepackage{amssymb}
\usepackage{amsmath}

\newcounter{bla}

\journal{Computer Physics Communications}

\usepackage{graphicx}% Include figure files
\usepackage[normalem]{ulem}
\usepackage{color}
\usepackage{dcolumn}% Align table columns on decimal point
\usepackage{bm}% bold math
\usepackage{float}

\begin{document}

\begin{frontmatter}

%% Title, authors and addresses

%% use the tnoteref command within \title for footnotes;
%% use the tnotetext command for the associated footnote;
%% use the fnref command within \author or \address for footnotes;
%% use the fntext command for the associated footnote;
%% use the corref command within \author for corresponding author footnotes;
%% use the cortext command for the associated footnote;
%% use the ead command for the email address,
%% and the form \ead[url] for the home page:
%%
%% \title{Title\tnoteref{label1}}
%% \tnotetext[label1]{}
%% \author{Name\corref{cor1}\fnref{label2}}
%% \ead{email address}
%% \ead[url]{home page}
%% \fntext[label2]{}
%% \cortext[cor1]{}
%% \address{Address\fnref{label3}}
%% \fntext[label3]{}

\title{A Time-spectral Approach to Numerical Weather Prediction}

%% use optional labels to link authors explicitly to addresses:
%% \author[label1,label2]{<author name>}
%% \address[label1]{<address>}
%% \address[label2]{<address>}

\author[a]{Jan Scheffel\corref{author}}
\author[a]{Kristoffer Lindvall}
\author[b]{Hiu Fai Yik}

\cortext[author] {Jan Scheffel.\\\textit{E-mail address:} jan.scheffel@ee.kth.se}
\address[a]{Department of Fusion Plasma Physics, School of Electrical Engineering \\
	KTH Royal Institute of Technology, Stockholm, Sweden}
\address[b]{Hong Kong University of Science and Technology, \\
 Clear Water Bay, Kowloon, Hong Kong  \\}

\begin{abstract}
%% Text of abstract
Finite difference methods are traditionally used for modelling the time domain in numerical weather prediction (NWP). Time-spectral solution is an attractive alternative for reasons of accuracy and efficiency and because time step limitations associated with causal, CFL-like critera are avoided.  In this work, the Lorenz 1984 chaotic equations are solved using the time-spectral algorithm GWRM. Comparisons of accuracy and efficiency are carried out for both explicit and implicit time-stepping algorithms. It is found that the efficiency of the GWRM compares well with these methods, in particular at high accuracy. For perturbative scenarios, the GWRM was found to be as much as four times faster than the finite difference methods. A primary reason is that the GWRM time intervals typically are two orders of magnitude larger than those of the finite difference methods. The GWRM has the additional advantage to produce analytical solutions in the form of Chebyshev series expansions. The results are encouraging for pursuing further studies, including spatial dependence, of the relevance of time-spectral methods for NWP modelling. 

\end{abstract}

\begin{keyword}
%% keywords here, in the form: keyword \sep keyword
Time-spectral; spectral; weighted residual methods; NWP; Chebyshev.

\end{keyword}

\end{frontmatter}

%%
%% Start line numbering here if you want
%%
% \linenumbers

% Computer program descriptions should contain the following
% PROGRAM SUMMARY.
\iffalse  % \iffalse ... \fi comments out large sections
{\bf PROGRAM SUMMARY/NEW VERSION PROGRAM SUMMARY}
  %Delete as appropriate.

\begin{small}
\noindent
{\em Program Title:}                                          \\
{\em Licensing provisions(please choose one): CC0 1.0/CC By 4.0/MIT/Apache-2.0/BSD 3-clause/BSD 2-clause/GPLv3/CC BY NC 3.0 }                                   \\
{\em Programming language:}                                   \\

{\em Supplementary material:}                                 \\
  % Fill in if necessary, otherwise leave out.

{\em Journal reference of previous version:}                  \\
 {\em Does the new version supersede the previous version?:}   \\
 {\em Reasons for the new version:}\\
 {\em Summary of revisions:}*\\
{\em Nature of problem(approx. 50-250 words):}\\
  %Describe the nature of the problem here. \\
{\em Solution method(approx. 50-250 words):}\\
  %Describe the method solution here.
{\em Additional comments including Restrictions and Unusual features (approx. 50-250 words):}\\
  %Provide any additional comments here.
   \\

* Items marked with an asterisk are only required for new versions
of programs previously published in the CPC Program Library.\\
\end{small}
\fi

%% main text
\section{Introduction}
\subsection{Lorenz 1984 Model}
The study of atmospheric dynamics is challenging due to its complex and chaotic nature. Lorenz \cite{Lorenz:1984atmosphere} proposed in 1984 a simplified model for representing Hadley circulation of air in the atmosphere:
\begin{align}
\frac{dX}{dt}&=-Y^2-Z^2-aX+aF\\
\frac{dY}{dt}&=XY-bXZ-Y+G\\
\frac{dZ}{dt}&=bXY+XZ-Z
\end{align}

The equations are, in his words, "the simplest model capable of representing an unmodified or modified Hadley circulation, determining its stability, and, if it is unstable, representing a stationary or migratory disturbance" \cite{Lorenz:1984atmosphere}. Lorenz uses the variable $X$ to represent the intensity of the symmetric globe-encircling westerly wind current, and also the poleward temperature gradient. The variables $Y$ and $Z$ represent the cosine and sine phase of a chain of superposed large-scale eddies. The parameters $a$, $b$, $F$ and $G$ may be chosen within certain bounds.

The model demonstrates chaotic behaviour for certain sets of parameters. In want of analytical solutions, initial-value problems are traditionally solved purely numerically by the use of finite steps in the temporal domain. The time steps of explicit time advance methods for general, space-dependent problems are restricted to small values through constraints such as the Courant-Friedrichs-Lewy (CFL) condition, and implicit schemes require time-consuming matrix operations at each time step. Semi-implicit methods allow large time steps and more efficient matrix inversions than those of implicit methods, but may feature limited accuracy. In this work we suggest an alternative, time-spectral approach for solution of equations (1)-(3). 

\subsection{Generalized Weighted Residual Method}
The Generalized Weighted Residual Method (GWRM) differs from traditional spectral methods for initial-value problems \cite{Gottlieb:1} in that also the time domain is treated spectrally \cite{Scheffel:GWRM1}. As a result the GWRM eliminates grid causality conditions such as the CFL condition, being associated with time stepping algorithms. Although the problems to be solved are assumed causal, the method is acausal in the sense that the time dependence is calculated by a global minimization procedure (the weighted residual formalism) acting on the time integrated problem. Recall that in standard WRM \cite{Fletcher:1}, initial value problems are transformed into a set of of coupled ordinary, linear or non-linear, differential equations for the time-dependent expansion coefficients. These are solved using finite differencing techniques. 

The GWRM enables semi-analytical solutions;  finite approximate solutions are obtained as analytical Chebyshev expansions. Not only temporal and spatial, but also a physical parameter domain may be treated spectrally, being of interest for carrying out scaling dependence in a single computation. Chebyshev polynomials are used for the spectral representation. These have several desirable qualities. They converge rapidly to the approximated function, they are real and can straightforwardly be converted to ordinary polynomials and vice versa, their minimax property guarantees that they are the most economical polynomial representation, they can be used for non-periodic boundary conditions (being problematic for Fourier representations) and they are particularly apt for representing boundary layers where their extrema are locally dense \cite{Scheffel:GWRM1}.

In standard text books on spectral methods for differential equations, time-spectral methods are usually touched upon only briefly and dismissed on the grounds that they are expensive \cite{Canuto:1,Boyd:1}. Historically, a number of authors have investigated various suggestions for and aspects of spectral methods in time. 
In 1979 a pseudo-spectral method, based on iterative calculation and an approximate factorization of the given equations, was suggested in \cite{Morchoisne:1}. Also, some early ideas were not developed further by Peyret and Taylor in \cite{Peyret:1}.

In 1986 and 1989, Tal-Ezer \cite{Tal-Ezer:1,Tal-Ezer:2},  proposed time-spectral methods for linear, periodic hyperbolic and parabolic equations, respectively, using a polynomial approximation of the evolution operator in a Chebyshev least square sense. Periodicity was assumed for the spatial domain through use of the Fourier spectral approximation.
The method extends the traditional ${\Delta}t=O(1/N{^2})$ stability criterion for explicit algorithms, where the space resolution parameter $N=O(1/{\Delta}x)$, to higher efficiency resulting in the stability condition ${\Delta}t=O(1/N)$). This approach to extend the time step in explicit methods was further studied in \cite{Kosloff:1}. The method is not widely used; a reason for this may be its complexity and its restriction to certain classes of problems. Later, Luo extended the method to more general boundary conditions and multiple spatial dimensions \cite{Luo:1}.

Ierley et al. \cite{Ierley:1} solved a a class of nonlinear parabolic partial differential equations with periodic boundary conditions using a Fourier representation in space and a Chebyshev representation in time. Similarly as for the GWRM, the Burger equation and other problems were solved with high resolution. Tang and Ma \cite{Tang:1} also used a spatial Fourier representation for solution of parabolic equations, but introduced Legendre Petrov-Galerkin methods for the temporal domain.

In 1994, Bar-Yoseph et al \cite{Zrahia:1,Bar-Yoseph:1} used space-time spectral element methods for solving one-dimensional nonlinear advection-diffusion problems and second order hyperbolic equations. Chebyshev polynomials were later employed in space-time least-squares spectral element methods \cite{Maerschalck:1}.

A theoretical analysis of Chebyshev solution expansion in time and one-dimensional space, for equal spectral orders, was given in \cite{Dutt:1}. The minimized residuals employed were however different from those of the GWRM.

More recently Dehghan and Taleei \cite{Dehghan:1} found solutions to the non-linear Schr\"{o}dinger equation, using a time-space pseudo-spectral method where the basis functions in time and space were constructed as a set of Lagrange interpolants.

Time-spectral methods feature high order accuracy in time. For implicit finite difference methods, deferred correction may provide high order temporal accuracy \cite{Gustafsson:1,Gustafsson:2}. A relatively recent approach to increase the temporal efficiency of finite difference methods is time-parallelization via the parareal algorithm \cite{Lions:1}. This method, however, features rather low parallel efficiency and improvements have been suggested, for example the use of spectral deferred corrections \cite{Emmett:1}.

An interesting Jacobian-free Newton-Krylov method for implicit time-spectral solution of the compressible Navier-Stokes equations has recently been put forth by Attar \cite{Attar:1}. 

A time-spectral method for periodic unsteady computations, using a Fourier representation in time, was suggested in \cite{Gopinath:1} and further developed in \cite{Sicot:1} and \cite{Luder:1}. A generalization to quasi-periodic problems was developed in \cite{Mavriplis:1}. 

In summary, although time-spectral methods have been explored in various forms by several authors during the last few decades, and were found to be highly accurate, the GWRM as described in \cite{Scheffel:GWRM1} has not been pursued. The present work contributes to the evaluation of this method.

The structure of the paper is as follows. In section 2 we briefly review the general GWRM formalism for solving a set of pde's but subsequently restrict us to a discussion of an optimized solution of the ode's (1)-(3). A major goal of this study is to evaluate possible advantages of the GWRM in relation to finite difference methods (FDM). Thus a comparative numerical study of convergence, accuracy, and efficiency for short time interval solutions are presented in section 3. We have selected the explicit fourth order Runge-Kutta method (RK4) to represent efficient FDM. Comparisons will also be made with the implicit second order Lobatto IIIA (trapezoidal rule) and fourth order Lobatto IIIC methods, the latter being particularly apt for stiff problems \cite{Brenan:Lobatto}. The chaotic character of the Lorenz equations is predominant in longer time calculations. In section 4 we determine long time accuracy and efficiency as well as the Liapunov exponent for the scenario studied here, employing both the RK4 and the GWRM. In section 5 the predictability of the GWRM is studied. We find that the efficiency of the GWRM compares well with and may exceed that of the RK4 method. A discussion follows in section 6 and conclusions are given in section 7.

\section{GWRM formalism}

The GWRM \cite{Scheffel:GWRM1} solves a system of initial-value partial differential equations
\begin{equation}
\frac{\partial\mathbf{u}}{\partial t}=D\mathbf{u}+f
\end{equation}
Here $D$ denotes a linear or nonlinear matrix differential operator that may depend on physical variables ($t$, $\mathbf{x}$, and $\mathbf{u}$) as well as on physical parameters (denoted $\mathbf{p}$), and $f$=${f}(t,\mathbf{x};\mathbf{p})$ is a known source or forcing term. An approximate solution ansatz is assumed as a truncated multivariate series of first kind Chebyshev polynomials $T$, which becomes
\begin{equation}
\mathbf{u}(t,x;p)=\sum_{q=0}^{Q}\sum_{r=0}^{R}\sum_{s=0}^{S}\mathbf{a}_{qrs}T_q(\tau)T_r(\xi)T_s(P)
\end{equation}

for the case of one spatial dimension $x$ and one parameter $p$. The coefficients $\mathbf{a}_{qrs}$ of this analytical expression are obtained from the Galerkin weighted residual method. For a single differential equation in $u$ it involves integration
\begin{equation}
\int_{t_0}^{t_1}\int_{{x_0}}^{{x_1}}\int_{{p_0}}^{{p_1}}RT_q(\tau)T_r(\xi)T_s(P)w_tw_xw_pdtd{x}d{p}=0
\end{equation}
over the entire computational domain and solution for all indices $q,r,s$. The following weight factors are used \cite{Mason:1}:
\begin{equation}
w_t=(1-\tau^2)^{\frac{1}{2}},
w_x=(1-\xi^2)^{\frac{1}{2}},
w_p=(1-P^2)^{\frac{1}{2}}
\end{equation}
$R$ is the residual of the approximation \cite{Fletcher:1}, defined as
\begin{equation}
R=u(t,{x};{p})-\left(u(t_0,{x};{p})+\int_{t_0}^{t}(Du+f)dt'\right)
\end{equation}
and $\tau$, $\xi$ and $P$ are given by ($z$ can be any of $t$, $x$ or $p$ and $z_0$ and $z_1$ represent interval boundaries.)
\begin{equation}
\tau=\frac{t-A_t}{B_t},
\xi=\frac{x-A_x}{B_x},
P=\frac{p-A_p}{B_p},
\end{equation}
\begin{equation}
A_z=\frac{z_1+z_0}{2},
B_z=\frac{z_1-z_0}{2}
\end{equation}

Carrying out the integrations in equation (6), the following system of algebraic equations results for the Chebyshev coefficients: 
\begin{equation}
	a_{qrs}=2\delta_{q0}b_{rs}+A_{qrs}+F_{qrs}
\end{equation}
for which $b_{rs}$ are the Chebyshev coefficients of the initial condition, and $A_{qrs}$ are obtained from the expansion
\begin{equation}
\int_{t_0}^{t}Dudt'=\sum_{q=0}^{Q}\sum_{r=0}^{R}\sum_{s=0}^{S}A_{qrs}T_q(\tau)T_r(\xi)T_s(P)
\end{equation}
showing that equation (11) is a linear/nonlinear equation in the parameters $a_{qrs}$ depending on the form of equation (4). The coefficients $F_{qrs}$ are obtained by a similar procedure to that in (12). Boundary conditions enter by replacing high-end $r$ indexed coefficients of (11) with corresponding spectral boundary equations. For the system of ode's (1)-(3), the forcing term and boundary conditions are not applicable and equations (11)-(12) are replaced by
\begin{equation}
	\mathbf{a}_{q}=2\delta_{q0}\mathbf{b}+\mathbf{A}_{q}
\end{equation}
\begin{equation}
\mathbf{b}=(X(0),Y(0),Z(0)),  ~~~ \int_{t_0}^{t}D\mathbf{u}dt'=\sum_{q=0}^{Q}\mathbf{A}_{q}T_q(\tau)
\end{equation}
where the array $\mathbf{a}_q$ contains the Chebyshev coefficients of all $X$, $Y$ and $Z$ variables.

To iteratively solve the system of non-linear algebraic equations for the Chebyshev coefficients in the GWRM, a semi-implicit root-solver (SIR) has been developed \cite{Scheffel:SIR}. This solver has better global convergence characteristics than Newton's method and avoids landing on local minima, as may occur when employing Newton's method with line-search. In GWRM applications, the handling of matrix equations in SIR for solution of equations (11) or (13) is the bottleneck both in terms of memory use and efficiency. In unoptimized cases, memory scales with the total number of Chebyshev coefficients to the second power, here as $(3(Q+1))^2$, and the number of operations scales to the third power, here $(3(Q+1))^3$. There exist, however, optimizing metods not touched upon here to reduce these scalings substantially. 

For the present ode's to be solved, the maximum Chebyshev order $Q$ will be 10 or less, so that the main factor governing efficiency is rather the rate of convergence, determined by the number of SIR iterations which, in turn, depend on the length of the time interval used. A single GWRM solution for temporal domains exceeding about two Lorenz time units can generally not be computed due to the complexity of the present equations and the spurious roots that are generated by the nonlinear algebraic terms obtained by transforming equations (1)-(3) to equation (13). The time domain is thus divided into a number of partially overlapping subintervals of typical length 0.5-2 Lorenz time units. End conditions at each time interval are used as initial conditions for the subsequent interval. The solutions are finally joined to constitute a piece-wise analytical solution for the entire time domain. It should be noted that the lengths of the time intervals exceed the length of finite difference time steps with orders of magnitude. In this work, GWRM time intervals are typically a factor 100 longer than the RK4 time steps for comparable accuracy. 

Furthermore, the GWRM code employed here uses an adaptive algorithm to control the length of each time interval in order to guarantee convergence. If the accuracy requirements are not fulfilled, a shorter time interval will be chosen and the coefficients will be recalculated. The amount of overlap between subsequent time domains can be arbitrarily chosen in order to provide optimized two-point contact and enhanced convergence. 

Generally, the convergence of iterative solvers like SIR are strongly dependent on the choice of starting conditions. An important property of the GWRM is that convergence is guaranteed as the time interval is decreased. This is because the solution then approaches the initial state, the coefficients of which are used as initial iterate for SIR. Moreover we will see that for simulations that are performed as perturbations of a "base run", use of the latter solution as initial state significantly reduces the number of required SIR iterations. The GWRM efficiency then exceeds that of RK4 and implicit methods by a wide margin.

\section{Convergence, accuracy, and efficiency}

We are interested in solving equations (1)-(3) accurately for long times up to some 100 time units. Consequently it is of interest to optimize the GWRM solution in individual time intervals. We may thus ask: is it preferable to use higher order approximations in long time intervals rather than to use lower order approximations in shorter intervals? To find out,  we solve equations (1)-(3) for $a = 0.25$, $b = 4.0$, $F = 8.0$ and $G = 1.0$, which parameters correspond to chaotic behaviour. The initial conditions used are well inside the attractor domain; $(X(0),Y(0),Z(0))=(0.96,-1.1,0.5)$. A corresponding GWRM solution for the time interval $[0,30]$ is shown in FIG. 1.

\begin{figure}[h!]
	\includegraphics[width=5.0in]{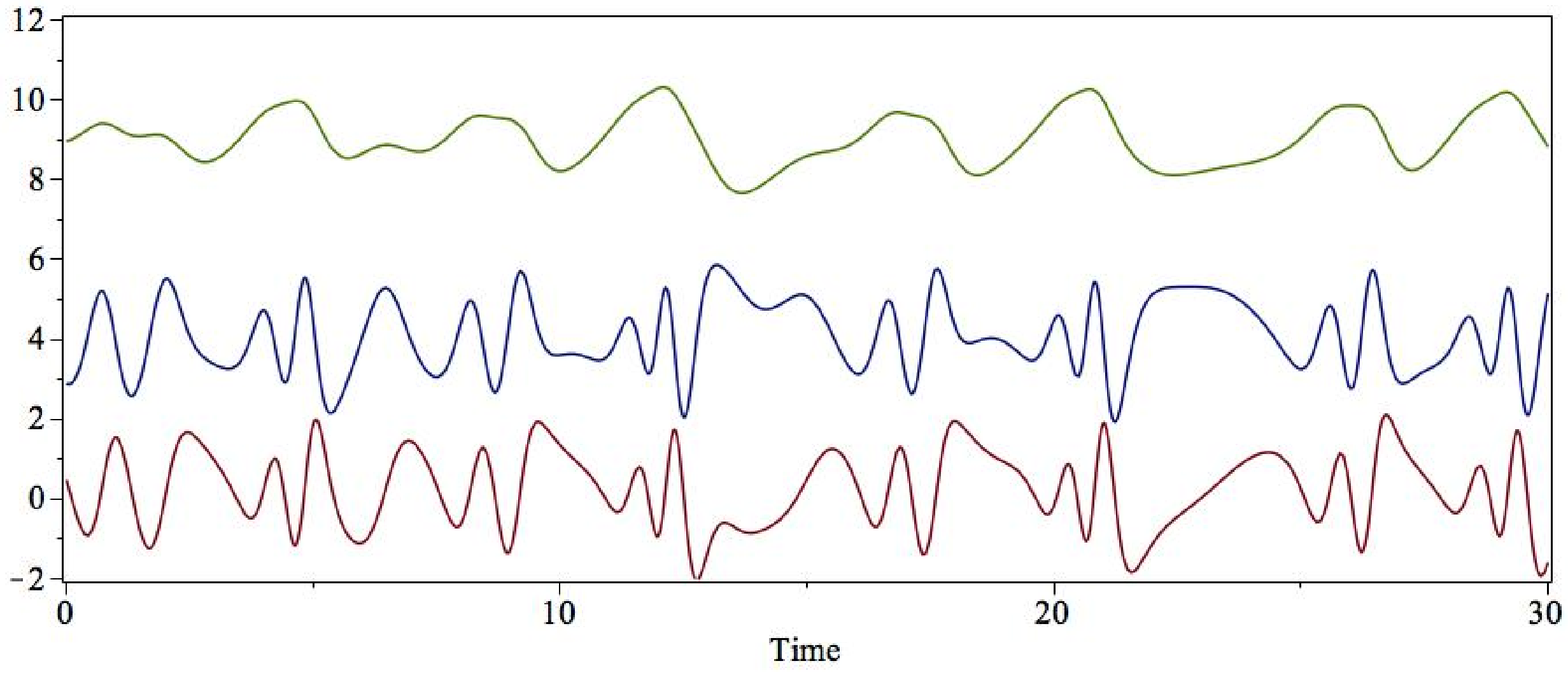}
	\caption{GWRM solution of Lorenz equations (1)-(3) for initial values $(X(0),Y(0),Z(0))=(0.96,-1.1,0.5)$ and parameters $a = 0.25$, $b = 4.0$, $F = 8.0$ and $G = 1.0$. From top to bottom; $X(t)+8, Y(t)+4, Z(t)$. The time domain was divided into $N=71$ subintervals during the time-adaptive computation.}
\end{figure}

In order to control GWRM accuracy we primarily need to consider the parameters used in SIR, the accuracy of which is essentially determined by three parameters. The first is the round-off error which, for the Maple computations of this paper, is globally set to $10^{-14}$. Second, a parameter $tol$ governs the mean absolute difference between successively iterated Chebyshev coefficients. Third, a parameter $\epsilon$ controls the numerical ratio between the highest and lowest two mode number Chebyshev coefficients (absolute values); this ratio is small for well resolved solutions. 

In TABLE~\ref{Table_maxinterval} approximate maximum time interval lengths, corresponding to convergent GWRM solutions, are shown for different Chebyshev orders $Q$. CPU times for each case are also given. We have here set $tol=10^{-5}$. SIR is run in standard mode, without sub-iterations; see \cite{Scheffel:SIR}. The convergence of Newton's method would be insufficient for some of these cases. When solving for longer time intervals than those of TABLE~\ref{Table_maxinterval}, the GWRM may still converge but to spurious solutions that are not causally related to the initial conditions assumed. For the same reason, values of $Q$ beyond 8 do not result in convergent solutions in longer time intervals. We can now answer the question posed above; it is usually more efficient to employ lower values of $Q$ in time subintervals of some optimized length than to use a higher value in a single interval with the same total length. 

\begin{table}[h!]
	\caption{Maximum time interval lengths allowing GWRM convergence, and corresponding CPU times in seconds.}
\begin{center}
	\begin{tabular}{|c|c|c|}\hline
		$Q$&Time interval&CPU time\\\hline
		$4$&$1.0$&0.031\\\hline
		$5$&$1.3$&0.046\\\hline
		$6$&$1.6$&0.047\\\hline
		$7$&$1.9$&0.047\\\hline
		$8$&$2.0$&0.062\\\hline
		$9$&$1.7$&0.062\\\hline
		$10$&$1.8$&0.078\\\hline
	\end{tabular}\label{Table_maxinterval}
\end{center}
\end{table}

We need also to consider accuracy. In order to determine the values required of the parameters $Q$, $tol$ and $\epsilon$ for a certain accuracy, comparisons with highly accurate solutions of (1)-(3) should be carried out. The solutions are obtained from Maple's built-in ode solver dsolve, for which we have set the relative error to $10^{-10}$. A series of computations have been made for the time interval $[0,2]$. The automatic time-adaptive algorithm, here using a time interval overlap of length $10^{-8}$ time units, checks for convergence. If convergence is not obtained, the time interval is halved. Every $M$th interval, where $M$ can be set (it is here 10), the code attempts to extend the time interval by a given factor, here 1.5. Obtained accuracy, in terms of maximum error of either of the $X$, $Y$, or $Z$ variables, are displayed in TABLE~\ref{Table_max_error_at_2}. Again we have set $tol=10^{-5}$. The number of time subintervals used is denoted by $N$.

\begin{table}[h!]
	\caption{Maximum GWRM absolute error of $X,Y,Z$ at $t=2$ for $tol=10^{-5}$.}
\begin{center}
	\begin{tabular}{|c|c|c|c|c|}\hline
		$\epsilon$&$Q$&$N$&CPU time&Error\\\hline
		$10^{-1}$&4&$7$&0.078&$7.9\times10^{-2}$\\\hline
		$10^{-1}$&6&$3$&0.062&$1.6\times10^{-2}$\\\hline
		$10^{-1}$&8&$3$&0.078&$5.6\times10^{-5}$\\\hline
		$10^{-1}$&10&$3$&0.109&$5.6\times10^{-5}$\\\hline
		$10^{-2}$&4&$12$&0.109&$1.9\times10^{-2}$\\\hline
		$10^{-2}$&6&$3$&0.078&$1.6\times10^{-2}$\\\hline
		$10^{-2}$&8&$3$&0.094&$1.1\times10^{-3}$\\\hline
		$10^{-2}$&10&$3$&0.109&$5.6\times10^{-5}$\\\hline
		$10^{-3}$&4&$28$&0.218&$1.7\times10^{-3}$\\\hline
		$10^{-3}$&6&$7$&0.125&$3.1\times10^{-4}$\\\hline
		$10^{-3}$&8&$3$&0.078&$1.1\times10^{-3}$\\\hline
		$10^{-3}$&10&$3$&0.110&$5.6\times10^{-5}$\\\hline
		$10^{-5}$&4&$176$&0.998&$1.2\times10^{-5}$\\\hline
		$10^{-5}$&6&$18$&0.250&$1.5\times10^{-5}$\\\hline
		$10^{-5}$&8&$7$&0.156&$1.2\times10^{-6}$\\\hline
		$10^{-5}$&10&$7$&0.250&$6.8\times10^{-7}$\\\hline
		$10^{-5}$&12&$3$&0.140&$2.9\times10^{-6}$\\\hline
	\end{tabular}\label{Table_max_error_at_2}
\end{center}
\end{table}

It is seen that, as far as accuracy is concerned, low $Q$ values are undesirable. In order to satisfy the $\epsilon$ criterion, the corresponding solutions need a large number of time subintervals and are thus costly. The solution error at the end of the computation ("Error" in TABLE~\ref{Table_max_error_at_2}) need be of order $10^{-3}$ for the computations of sections IV-V of this study. Thus $Q$-values in the range of 8-10 are found to be optimal with respect to both accuracy and efficiency. The root solver SIR is here best run in Newton mode to reduce the number of iterations required. The initial time interval is set to $[0,2/N_t]$ with $N_t=3$.

Next, we compare with the RK4 method. The single parameter that controls accuracy is the step length parameter $h$. In TABLE~\ref{Table_RK4_at_20} the maximum absolute error of $X,Y,Z$ at $t=20$ is computed for various step lengths. Corresponding results for the GWRM are displayed for different values of $\epsilon$ and $tol$ (with $N_t=50$) in TABLE~\ref{Table_GWRM_fixed_tol_at_20} and TABLE~\ref{Table_GWRM_fixed_epsilon_at_20} respectively. It is seen that for low accuracy RK4 is somewhat more efficient than GWRM. For higher accuracy (better than $10^{-3}$) the GWRM is more efficient than RK4. This is particularly pronounced when the time overlap ("handshaking") procedure in GWRM is omitted. It should be noted that similar results as in TABLE~\ref{Table_GWRM_fixed_tol_at_20} and TABLE~\ref{Table_GWRM_fixed_epsilon_at_20} are obtained when the initial conditions $(X(0),Y(0),Z(0))$ are varied. The results are thus representative. 

RK4 is an explicit method. Are implicit algorithms like the second-order Trapezoid or the fourth-order Lobatto IIIC methods possibly more efficient? In TABLES \ref{Table_Trapezoid_at_20} and \ref{Table_LobattoIIIC_at_20} we show results for these methods for the same step lengths that were used for RK4 in TABLE III. It is seen that both methods are more costly than RK4 and GWRM for the same accuracy. The Lobatto IIIC method is more than an order of magnitude slower.

\begin{table}[h!]
	\caption{RK4 method maximum absolute error of $X,Y,Z$ at $t=20$.}
\begin{center}
	\begin{tabular}{|c|c|c|c|}\hline
		$h$&Steps&CPU time&Error\\\hline
		$0.1$&200&$0.21$&$1.3\times10^{0}$\\\hline
		$0.05$&400&$0.41$&$1.5\times10^{-1}$\\\hline
		$0.02$&1000&$1.01$&$4.8\times10^{-3}$\\\hline
		$0.01$&2000&$1.96$&$3.2\times10^{-4}$\\\hline
		$0.005$&4000&$3.91$&$5.7\times10^{-5}$\\\hline	
	\end{tabular}\label{Table_RK4_at_20}
\end{center}
\end{table}
\begin{table}[h!]
	\caption{GWRM maximum absolute error of $X,Y,Z$ at $t=20$ for $tol=10^{-5}$.}
\begin{center}
	\begin{tabular}{|c|c|c|c|c|}\hline
		$\epsilon$&$Q$&$N$&CPU time&Error\\\hline
		$10^{-1}$&6&$38$&0.48&$9.1\times10^{-1}$\\\hline
		$10^{-1}$&8&$37$&0.80&$1.8\times10^{-1}$\\\hline
		$10^{-1}$&10&$37$&1.12&$5.5\times10^{-3}$\\\hline
		$10^{-2}$&6&$51$&0.64&$1.3\times10^{-1}$\\\hline
		$10^{-2}$&8&$38$&0.73&$2.8\times10^{-2}$\\\hline
		$10^{-2}$&10&$37$&1.14&$5.5\times10^{-3}$\\\hline		
		$10^{-3}$&6&$72$&0.84&$7.2\times10^{-2}$\\\hline
		$10^{-3}$&8&$50$&0.91&$2.3\times10^{-2}$\\\hline
		$10^{-3}$&10&$38$&1.05&$2.4\times10^{-3}$\\\hline		
		$10^{-5}$&6&$185$&1.84&$1.2\times10^{-3}$\\\hline
		$10^{-5}$&8&$83$&1.42&$3.8\times10^{-4}$\\\hline
		$10^{-5}$&10&$55$&1.44&$5.6\times10^{-5}$\\\hline
	\end{tabular}\label{Table_GWRM_fixed_tol_at_20}
\end{center}
\end{table}
\begin{table}[H]
	\caption{GWRM maximum absolute error of $X,Y,Z$ at $t=20$ for $\epsilon=10^{-5}$.}
\begin{center}
	\begin{tabular}{|c|c|c|c|c|}\hline
		$tol$&$Q$&$N$&CPU time&Error\\\hline
		$10^{-1}$&6&$173$&1.20&$1.1\times10^{0}$\\\hline
		$10^{-1}$&8&$95$&1.12&$1.5\times10^{0}$\\\hline
		$10^{-1}$&10&$58$&1.08&$1.9\times10^{0}$\\\hline
		$10^{-2}$&6&$200$&1.65&$4.8\times10^{-2}$\\\hline
		$10^{-2}$&8&$84$&1.22&$2.0\times10^{-1}$\\\hline
		$10^{-2}$&10&$55$&1.17&$2.5\times10^{-2}$\\\hline
		$10^{-3}$&6&$185$&1.73&$1.8\times10^{-3}$\\\hline
		$10^{-3}$&8&$83$&1.28&$4.6\times10^{-4}$\\\hline
		$10^{-3}$&10&$55$&1.30&$4.2\times10^{-4}$\\\hline
		$10^{-5}$&6&$185$&1.84&$1.2\times10^{-3}$\\\hline
		$10^{-5}$&8&$83$&1.42&$3.8\times10^{-4}$\\\hline
		$10^{-5}$&10&$55$&1.44&$5.6\times10^{-5}$\\\hline
	\end{tabular}\label{Table_GWRM_fixed_epsilon_at_20}
\end{center}
\end{table}
\begin{table}[H]
	\caption{Trapezoid method maximum absolute error of $X,Y,Z$ at $t=20$.}
\begin{center}
	\begin{tabular}{|c|c|c|c|}\hline
		$h$&Steps&CPU time&Error\\\hline
		$0.1$&200&$0.84$&$1.1\times10^{0}$\\\hline
		$0.05$&400&$1.44$&$6.3\times10^{-1}$\\\hline
		$0.02$&1000&$3.33$&$8.0\times10^{-1}$\\\hline
		$0.01$&2000&$6.89$&$2.1\times10^{-1}$\\\hline
		$0.005$&4000&$12.6$&$8.0\times10^{-2}$\\\hline
		\end{tabular}\label{Table_Trapezoid_at_20}
\end{center}
\end{table}

\begin{table}[H]
	\caption{Lobatto IIIC method maximum absolute error of $X,Y,Z$ at $t=20$.}
\begin{center}
	\begin{tabular}{|c|c|c|c|}\hline
		$h$&Steps&CPU time&Error\\\hline
		$0.1$&200&$4.99$&$5.8\times10^{0}$\\\hline
		$0.05$&400&$9.32$&$4.8\times10^{-2}$\\\hline
		$0.02$&1000&$20.9$&$1.3\times10^{-3}$\\\hline
		$0.01$&2000&$40.4$&$4.6\times10^{-5}$\\\hline
		$0.005$&4000&$78.0$&$3.5\times10^{-5}$\\\hline	
	\end{tabular}\label{Table_LobattoIIIC_at_20}
\end{center}
\end{table}

\section{Long time behaviour}
%\subsection{Accuracy and Convergence}
To examine the accuracy of numerical solutions in absence of analytical solutions, a common practice is to study the convergence of different solutions, controlled by some parameter, towards a high accuracy solution. This approach will now be used in a comparative study of the accuracy of RK4, Lobatto IIIC and the GWRM for the Lorenz 1984 problem in the time interval $[0,30]$. Chaotic systems like the Lorenz 1984 equations are characterized by the fact that two initially adjacent states (with separation $E_{initial}$) will, even in absence of numerical errors, deviate at a rate 
\begin{equation}
E(t)=E_{initial}e^{\lambda t}
\end{equation}
where $\lambda$ is the Lyapunov exponent \cite{Lyapunov:1892thesis}. 

Our numerical results show that $\ln E$ typically grows rapidly when $t$ is small and then linearly when $t$ is large. This suggests an approximate dependence of the form $E(t)=F(t)+Ce^{At}$ where $F(t)$ is some non-exponential function, playing a dominant role when $t$ is small. For large $t$, the exponential factor takes over and the logarithmic graph is linear. 

In light of the above, we have found the following numerical procedure to be useful. For each of the numerical methods, a set of 100 solutions have been produced employing a set of selected accuracies. Each solution is compared with a high accuracy solution that uses exactly the same initial conditions, and the deviation as a function of time has been computed as described below. To optimize statistics, the attractor space is scanned by letting the end state of each high accuracy solution constitute the initial state for the new pair of solutions to be computed. The graphs $\ln(E)$ vs $t$ are subsequently generated by taking the arithmetic mean of the calculations.

The expression
\begin{equation}
\ln(E)=E_0+At
\end{equation}
is fitted to the graphs over the entire temporal domain. We find that the slope $A$ is quite independent of the choice of method or accuracy, which suggests that it is determined by the Lorenz system (1)-(3) itself indicating that $A=\lambda$. The accuracy of the method can thus be solely estimated from $E_0$. Since the $\ln E(t)$ graph is similar for each variable $X,Y,Z$, as can be seen from FIG. 2, we will mainly focus our discussion on the $X$ dimension.  

%\paragraph{GWRM}
\begin{figure}[h!]
\center
	\includegraphics[width=2.3in]{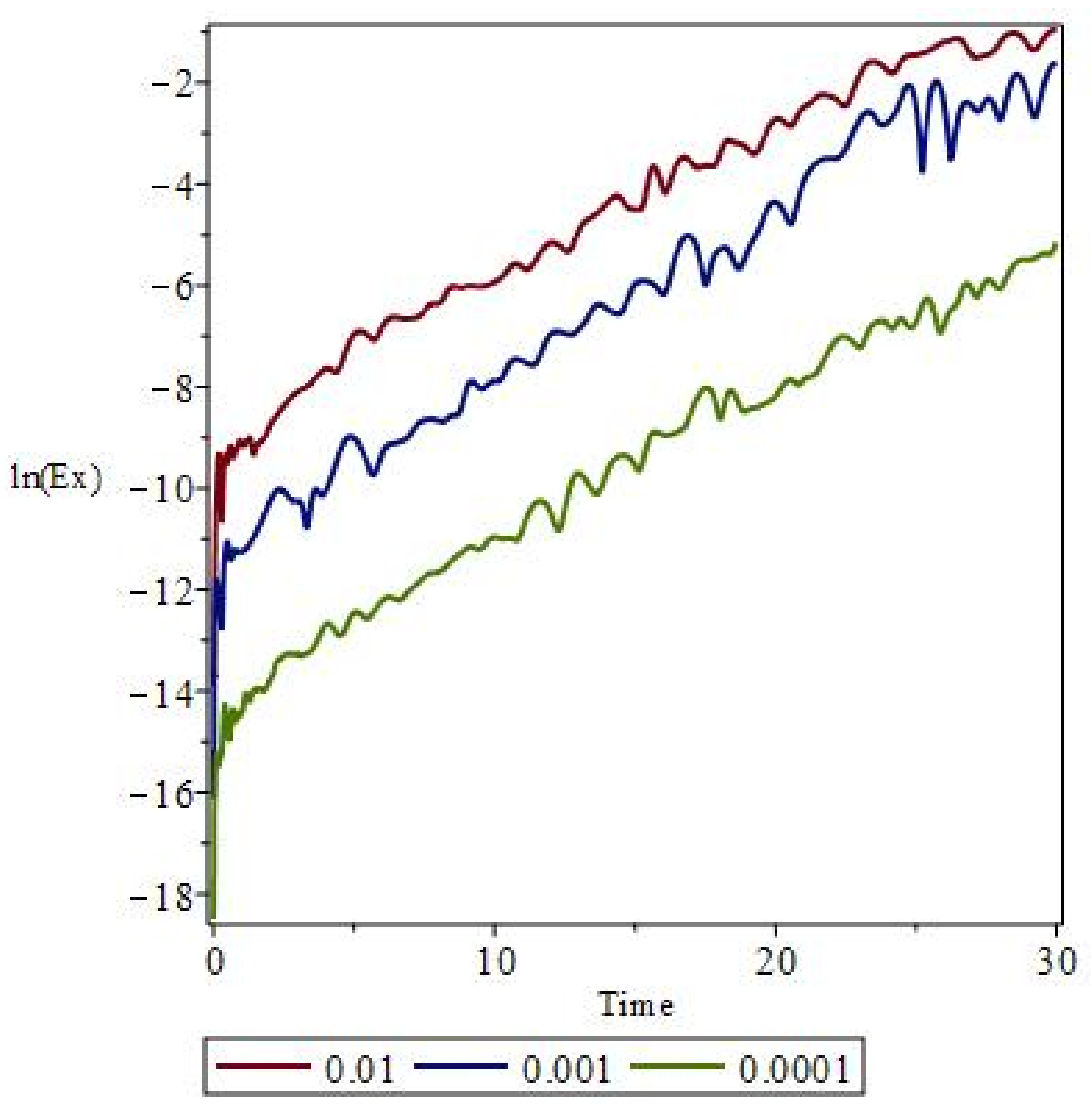}
	\includegraphics[width=2.3in]{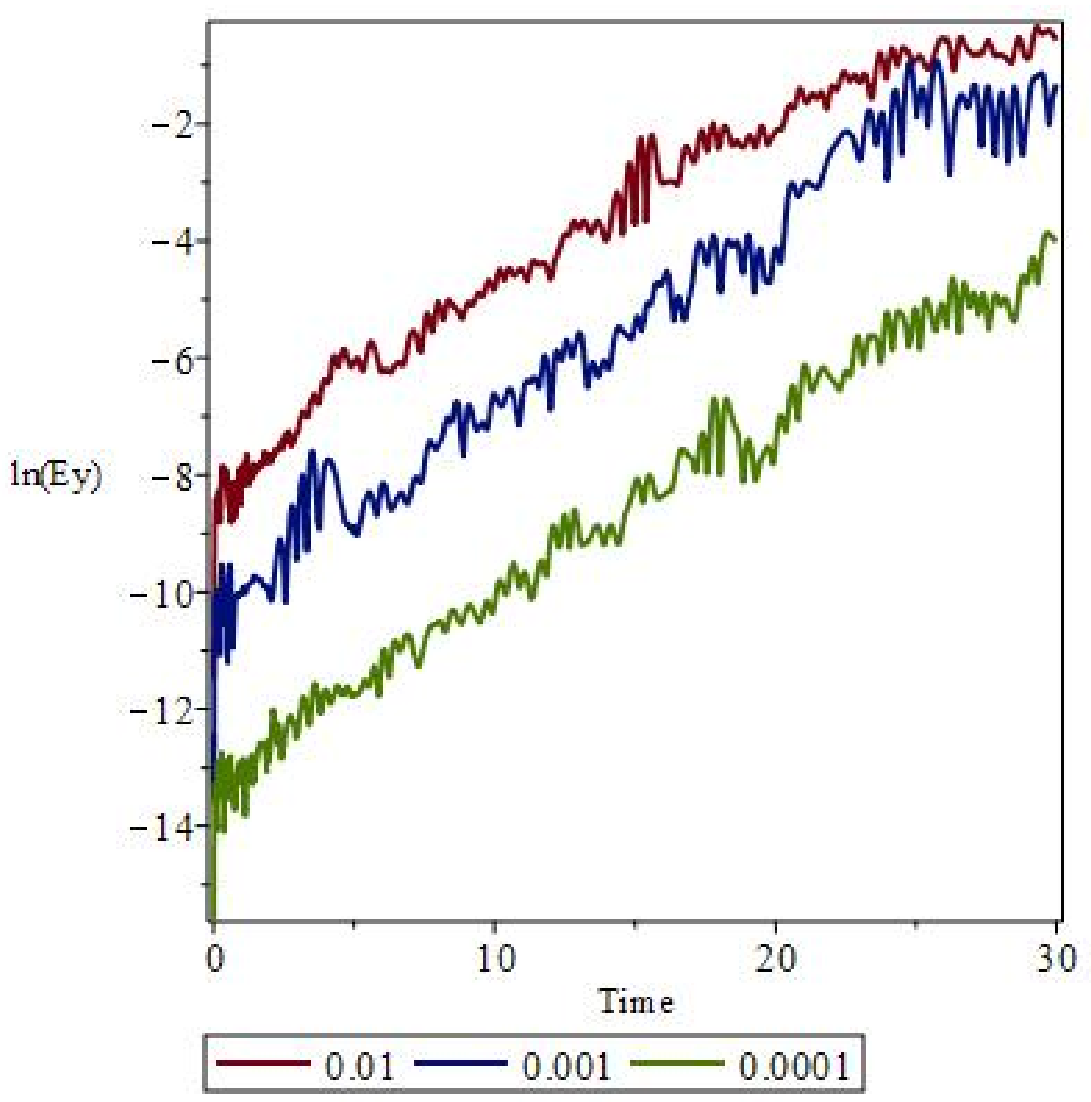}
	\includegraphics[width=2.3in]{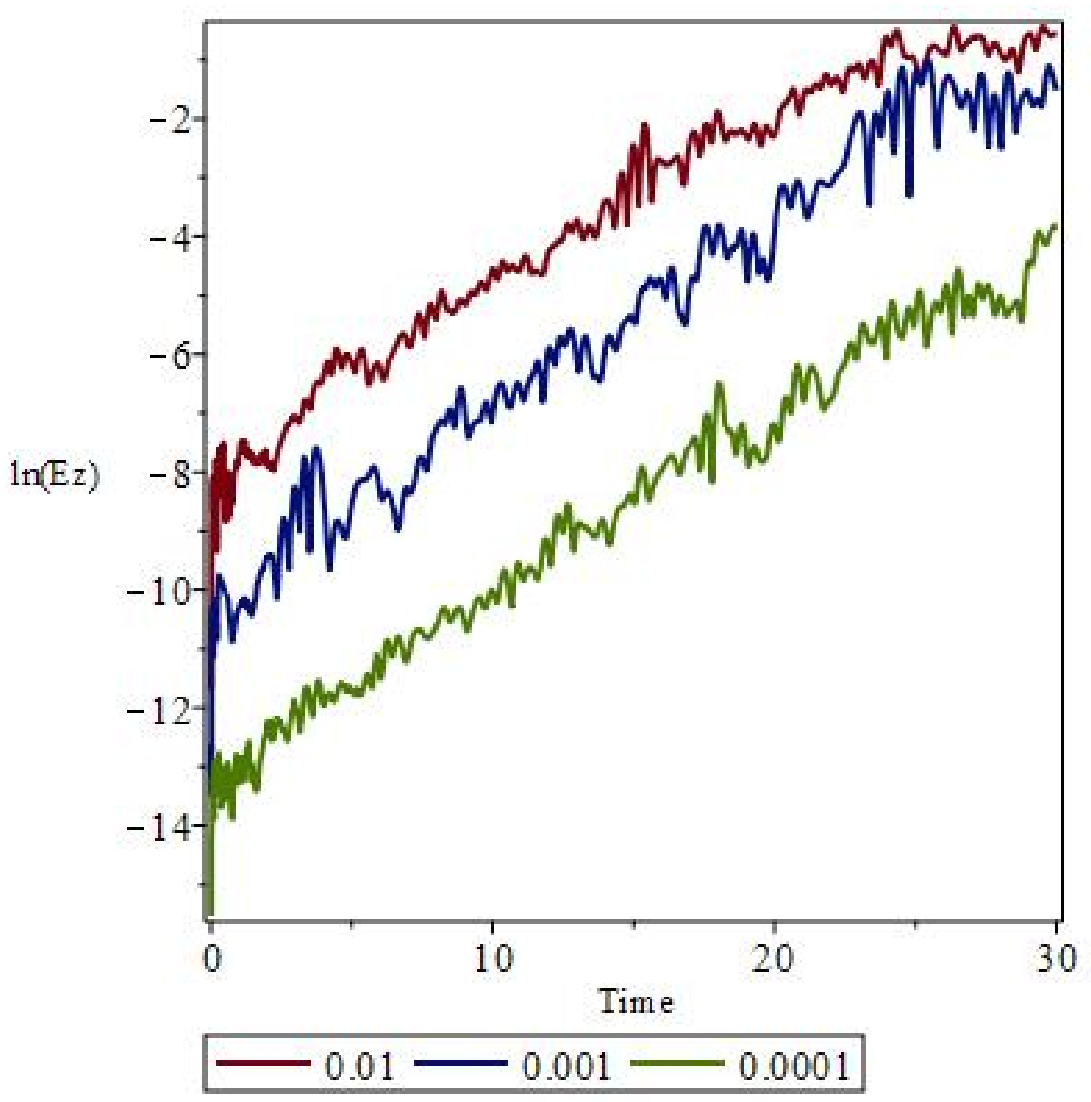}
	\caption{GWRM: $\ln E_{X,\epsilon,tol}$, $\ln E_{Y,\epsilon,tol}$ and $\ln E_{Z,\epsilon,tol}$ vs $t$ for $tol=10^{-5}$ and different $\epsilon$.}
\end{figure}
\begin{figure}[h!]
\center
	\includegraphics[width=2.3in]{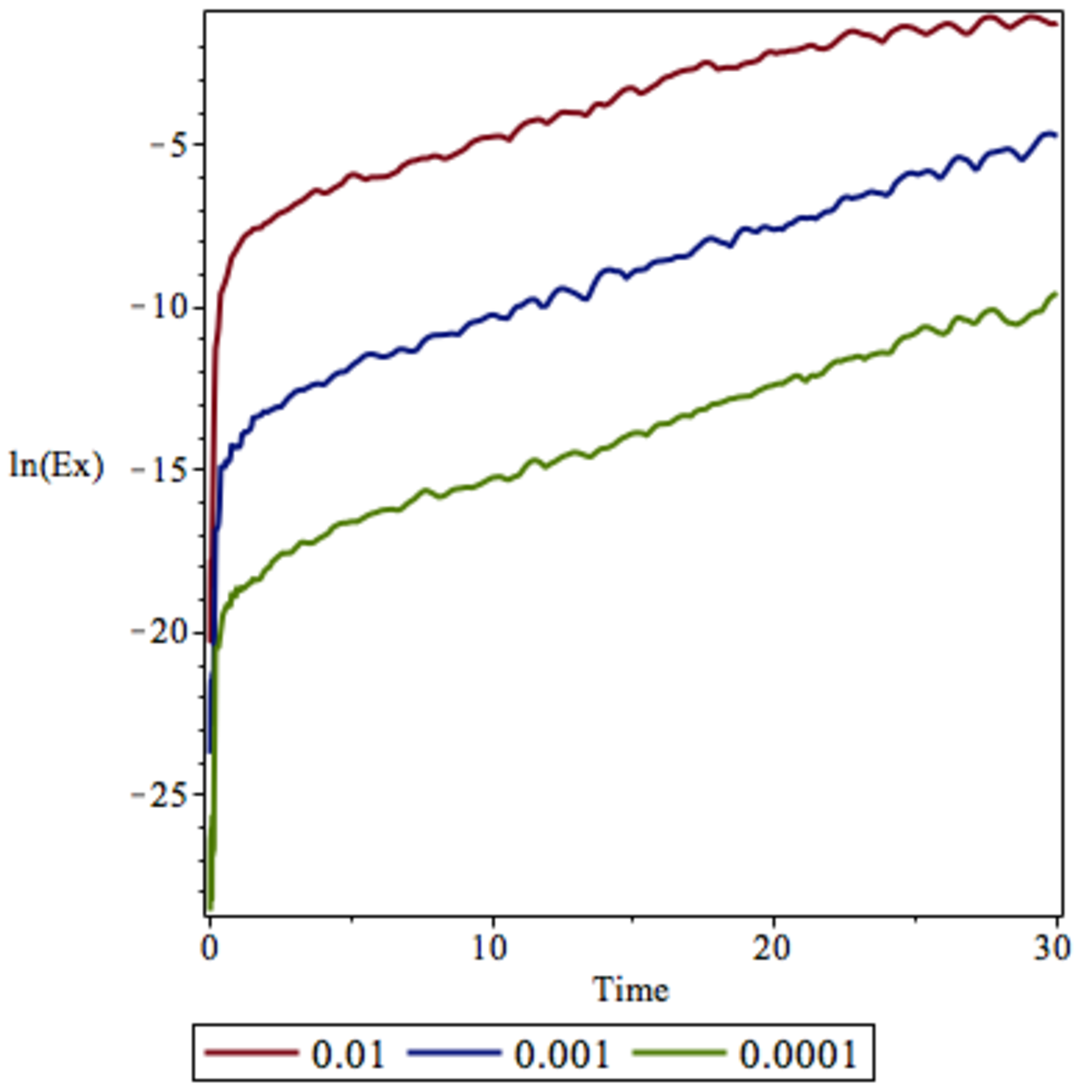}
	\caption{GWRM: $\ln E_{X,\epsilon,tol}$ vs $t$ for $\epsilon=10^{-5}$ and different $tol$.}
	\end{figure}
Two factors control the accuracy of the GWRM method: the SIR parameters $tol$ and $\epsilon$.
The deviation for each pair of GWRM computations is calculated by
\begin{equation}
%E_{$\epsilon,tol$}(t)=|{u}_{$\epsilon,tol$}-\mathbf{u}_{10^{-5},10^{-5}}|
E_{k,\epsilon,tol}(t)=|{u_{k}(t)}_{\epsilon,tol}-{u_{k}(t)}_{10^{-5},10^{-5}}|
\end{equation}
where $u_{k}$ denote the components of the vector $(X,Y,Z)$. Plots of $ln(E_{k,\epsilon,tol}(t))$ for various $\epsilon$ are given in FIG. 2 and a plot showing the deviations due to different $tol$ is shown in FIG. 3. Least square fitted parameters of (16) are listed in TABLE~\ref{Table_errorfit_GWRM}.
\begin{table}[h!]
	\caption{GWRM: Fitted parameters of Equation(17).}
\begin{center}
	\begin{tabular}{|c|c|c|c|}\hline
		$\varepsilon$&tol&$A$&$E_0$\\\hline
		$1\times 10^{-5}$&$1\times 10^{-4}$&0.27&-17.9\\\hline
		$1\times 10^{-5}$&$1\times 10^{-3}$&0.27&-12.9\\\hline
		$1\times 10^{-5}$&$1\times 10^{-2}$&0.27&-7.4\\\hline
		$1\times 10^{-4}$&$1\times 10^{-5}$&0.27&-14.5\\\hline
		$1\times 10^{-3}$&$1\times 10^{-5}$&0.27&-11.4\\\hline
		$1\times 10^{-2}$&$1\times 10^{-5}$&0.27&-9.1\\\hline	
	\end{tabular}\label{Table_errorfit_GWRM}
\end{center}
\end{table}

%\paragraph{FDM}
\begin{figure}[h!]
\center
	\includegraphics[width=2.3in]{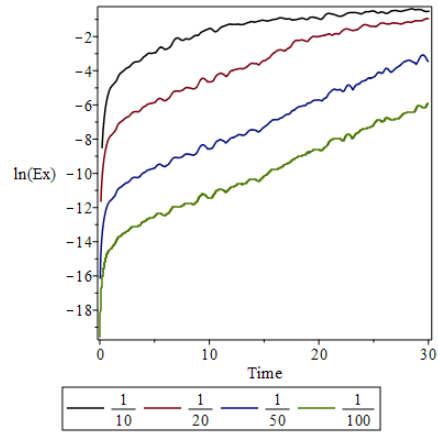}
	\caption{RK4: $\ln E_{X,h}$ vs $t$ with a variation of $h$.}
\end{figure}
\begin{figure}[h!]
\center
	\includegraphics[width=2.3in]{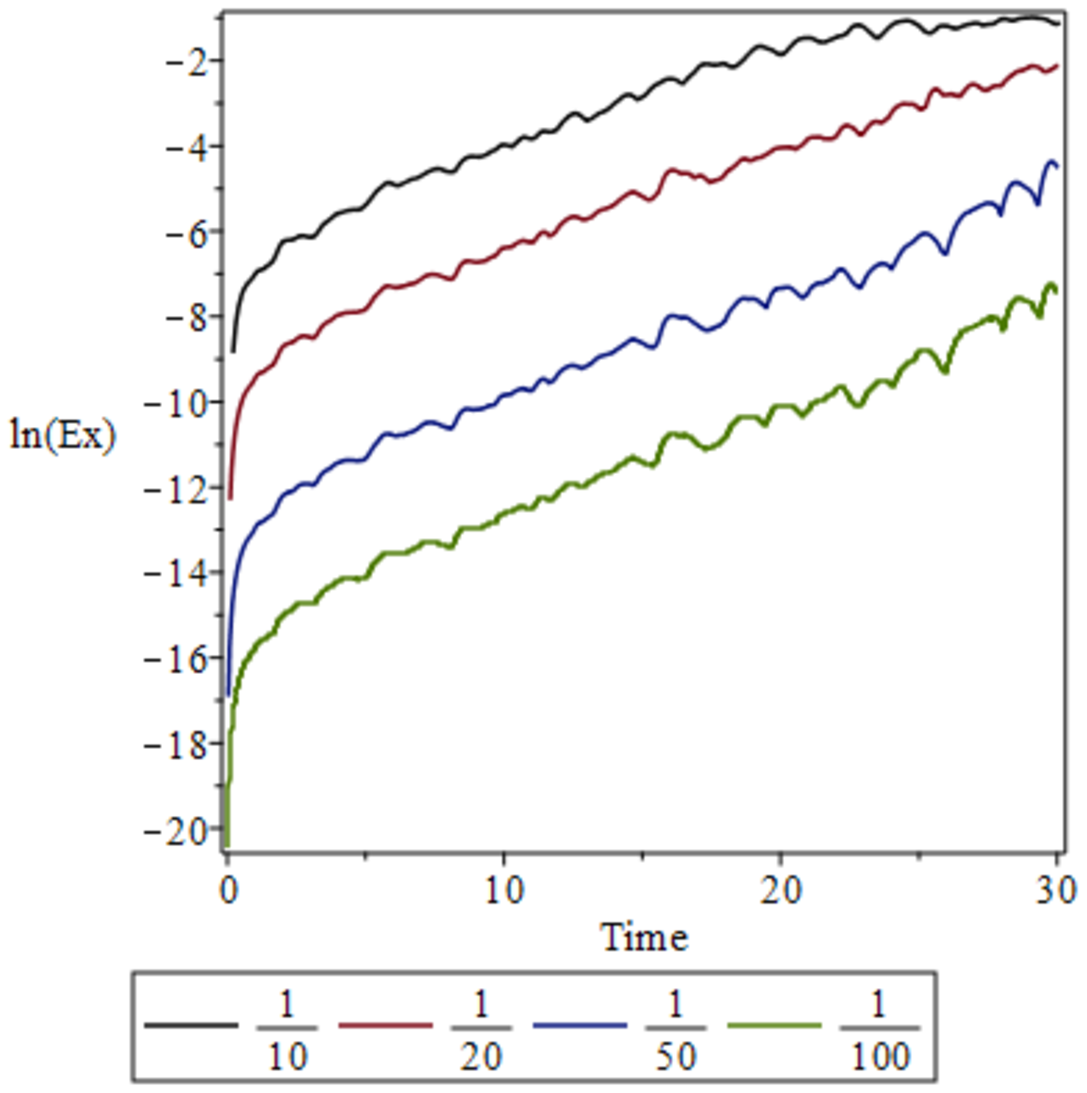}
	\caption{Lobatto IIIC: $\ln E_{X,h}$ vs $t$ with a variation of $h$.}
\end{figure}
The explicit classical RK4 and implicit Lobatto IIIC finite difference methods are chosen for comparisons.
For FDM, the controlling factor of accuracy is the time step $h$. The deviation is thus calculated as
\begin{equation}
E_{k,h}(t)=|{u_{k}(t)}_{h}-{u_{k}(t)}_{5\times 10^{-3}}|
\end{equation}
Deviations $E(t)$ for different time step lengths are plotted in FIG:s 4 and 5 respectively and the fitted parameters are listed in TABLE~\ref{Table_errorfit_RK4} and TABLE~\ref{Table_errorfit_LobattoIIIC}.

\begin{table}[h!]
	\begin{center}
		\caption{RK4: fitted parameters of Equation(17).}
		\begin{tabular}{|c|c|c|}\hline
			$h$&$A$&$E_0$\\\hline
			$1\times 10^{-1}$&0.27&-4.1\\\hline
			$5\times 10^{-2}$&0.27&-6.9\\\hline
			$2\times 10^{-2}$&0.27&-10.8\\\hline
			$1\times 10^{-2}$&0.27&-13.6\\\hline
	\end{tabular}\label{Table_errorfit_RK4}
	\end{center}
\end{table}
\begin{table}[h!]
	\begin{center}
		\caption{Lobatto IIIC: fitted Value of Equation(17).}
		\begin{tabular}{|c|c|c|}\hline
			$h$&$A$&$E_0$\\\hline
			$1\times 10^{-1}$&0.27&-6.8\\\hline
			$5\times 10^{-2}$&0.27&-9.1\\\hline
			$2\times 10^{-2}$&0.27&-12.8\\\hline
			$1\times 10^{-2}$&0.27&-15.4\\\hline
	\end{tabular}\label{Table_errorfit_LobattoIIIC}
	\end{center}
\end{table}
%\subsection{Efficiency}
The accuracy in terms of $E(0)$ of the fourth order FDM, the RK4 and Lobatto IIIC methods, is found to depend strongly on the step size; approximately $E(0)\propto h^{3.8}$. The accuracy scalings of the GWRM are $E_0\propto \text{$\epsilon$}^{1.2}$ for $tol=10^{-5}$ and $E_0\propto \text{$tol$}^{2.3}$ for $\epsilon=10^{-5}$; see TABLE XI. Comparing TABLES \ref{Table_errorfit_GWRM} and \ref{Table_errorfit_RK4} with run times in TABLES \ref{Table_RK4_at_20}, \ref{Table_GWRM_fixed_tol_at_20} and \ref{Table_GWRM_fixed_epsilon_at_20} it is again seen that the GWRM is more efficient at higher accuracies than the finite difference methods. 

\begin{table}[h!]
\center
\caption{Accuracy comparison.}
	\begin{tabular}{|c|c|c|c|c|c|}\hline
		Method&$\epsilon$&$tol$&$h$&$E_0$\\\hline
		RK4&&&$1\times 10^{-1}$&-4.1\\\hline
		&&&$5\times 10^{-2}$&-6.9\\\hline
		&&&$2\times 10^{-2}$&-10.8\\\hline
		&&&$1\times 10^{-2}$&-13.6\\\hline
		Lobatto IIIC&&&$1\times 10^{-1}$&-6.8\\\hline
		&&&$5\times 10^{-2}$&-9.1\\\hline
		&&&$2\times 10^{-2}$&-12.8\\\hline
		&&&$1\times 10^{-2}$&-15.4\\\hline
		GWRM&$1\times 10^{-2}$&$1\times 10^{-5}$&&-9.1\\\hline
		&$1\times 10^{-3}$&$1\times 10^{-5}$&&-11.4\\\hline
		&$1\times 10^{-4}$&$1\times 10^{-5}$&&-14.5\\\hline
		&$1\times 10^{-5}$&$1\times 10^{-2}$&&-7.4\\\hline
		&$1\times 10^{-5}$&$1\times 10^{-3}$&&-12.9\\\hline
		&$1\times 10^{-5}$&$1\times 10^{-4}$&&-17.9\\\hline
	\end{tabular}\label{Table_efficiency}
\end{table}

%\subsection{Perturbed Simulations}
In weather forecasting, running a number of simulations with slightly perturbed initial conditions is a common practice to enhance predictability. The GWRM is particularly well suited for these tasks, since the Chebyshev coefficients for the "base" run solution may be used as good initial iterates in SIR when computing the perturbed scenarios. This substantially reduces the number of iterations and the CPU time. In the next section predictability is studied in a comparison between the GWRM and the RK4 methods.  

\section{Predictability}

\noindent The Lorenz equations, and many weather related phenomena, are inherently chaotic. In other words, it is improbable that a long-term solution is representative if the initial condition varies slightly from the "exact" value. The error will eventually saturate at some level which implies that a random initial condition can be chosen without affecting the error growth, thus predictability is lost. The characteristic time $T$ when a perturbed solution diverges significantly from the true solution is thus of importance. It is in the time interval $[0,T]$ that predictions are meaningful.

It is of interest to compare the GWRM and RK4 method with regards to predictability. Our analysis goes as follows: a solution $\mathbf{u_{1}}$ is obtained for $t=50$, where $\mathbf{u}$ again denotes the vector $(X(t),Y(t),Z(t))$, and $"1"$ specifies a base run of the Lorenz equations with initial conditions  $\mathbf{V}=(X(0),Y(0),Z(0))$. Next we perturb the initial conditions slightly; $\mathbf{V}'=\mathbf{V}+\mathbf{\delta}$, where the three components of the vector $\mathbf{\delta}$ are randomly distributed so that $0<\delta<0.01$. The simulation is subsequently run again to obtain a perturbed solution $\mathbf{u}'_{n}$. The deviations $e_{k,n}=(u'_{k,n}-u_{k,1})^2$ between the two simulations are then calculated for $k=1,2,3$. The same procedure is used for all perturbed scenarios, and then summed over the scenarios in order to obtain an error growth $E_{k}(t)=\frac{1}{N}{\sum}(u'_{k,n}-u_{k,1})^2, ~n=2..N$, where $N$ is the number of scenarios. Both RK4 and GWRM undergo the same analysis with comparable accuracies, which is chosen as the average accuracy of the three variables X, Y and Z, whereby the CPU times used are collected in TABLE XII. 

The GWRM is found to be more than four times as efficient as RK4. As was shown in section III, this is partly due to the high efficiency of the GWRM for high accuracy, but mainly due to that for these perturbative runs, the initial iterates used by SIR are closer to the solution thus reducing the number of iterations needed.

\begin{table}[H]
\center
\caption{Error growth comparison between GWRM and RK4.}
	\begin{tabular}{|c|c|c|c|c|c|}\hline
		Numerical Method & Av. acc. & Steps & $\epsilon$ & \# Runs & CPU time [min] \\\hline
		GWRM & 5.6e-6 & & $10^{-5}$ & 250 & 17.9 \\\hline
		RK4 & 5.9e-6 & 15000 & & 250 & 83.5 \\\hline
	\end{tabular}\label{Table_errorgrowth}
%\label{tab:a}
\end{table}

\begin{figure}[H]
\center
	\includegraphics[width=2.3in]{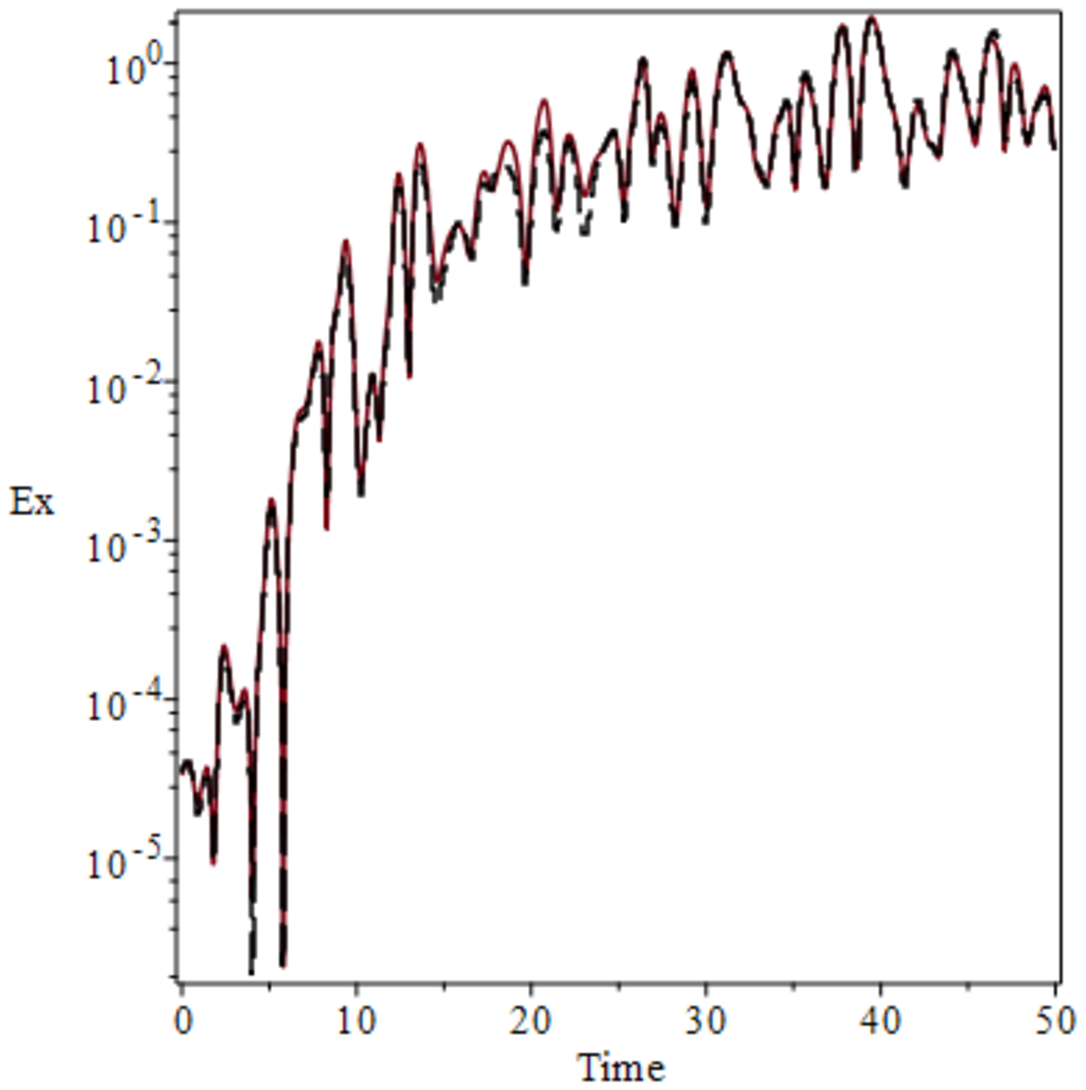}
	\includegraphics[width=2.3in]{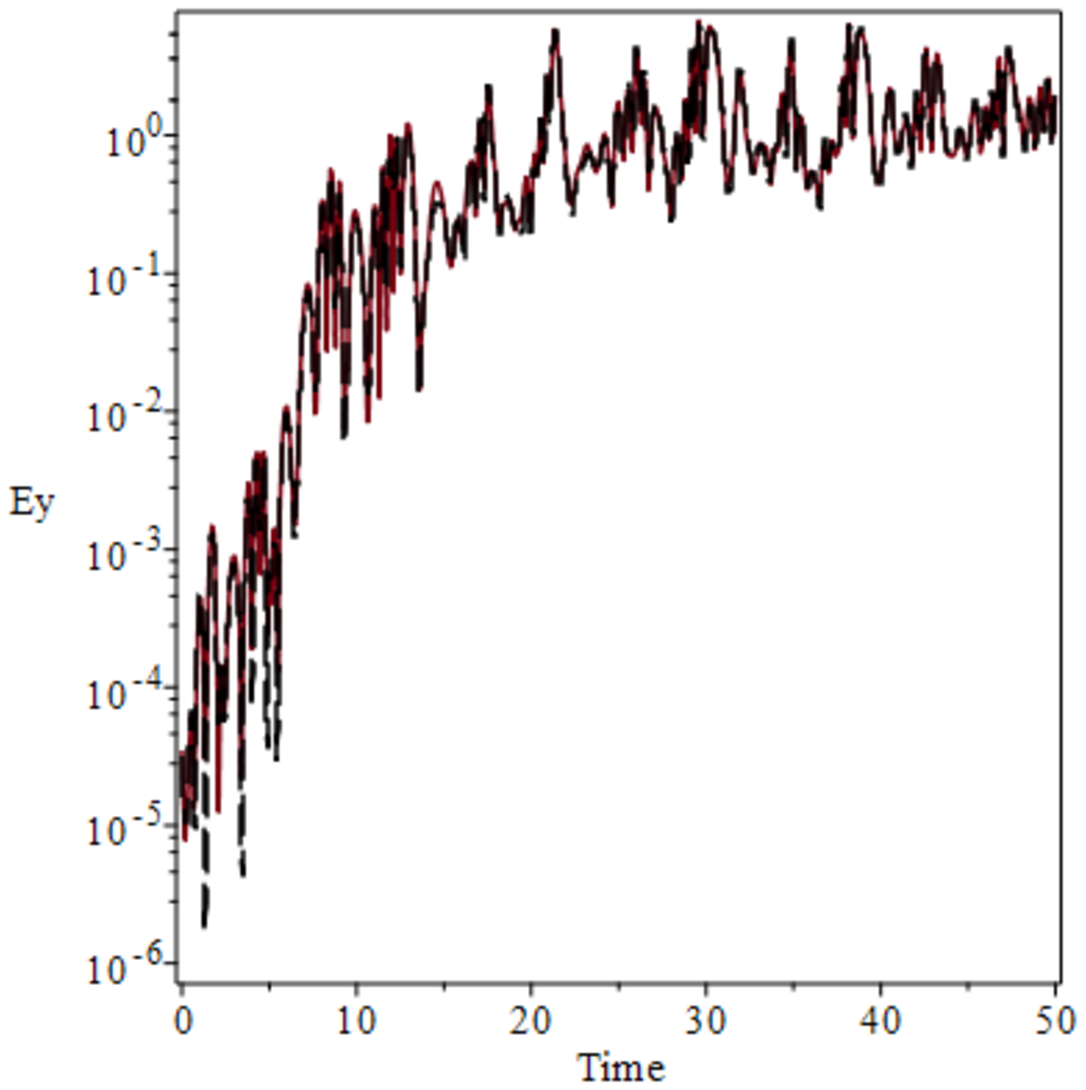}
	\includegraphics[width=2.3in]{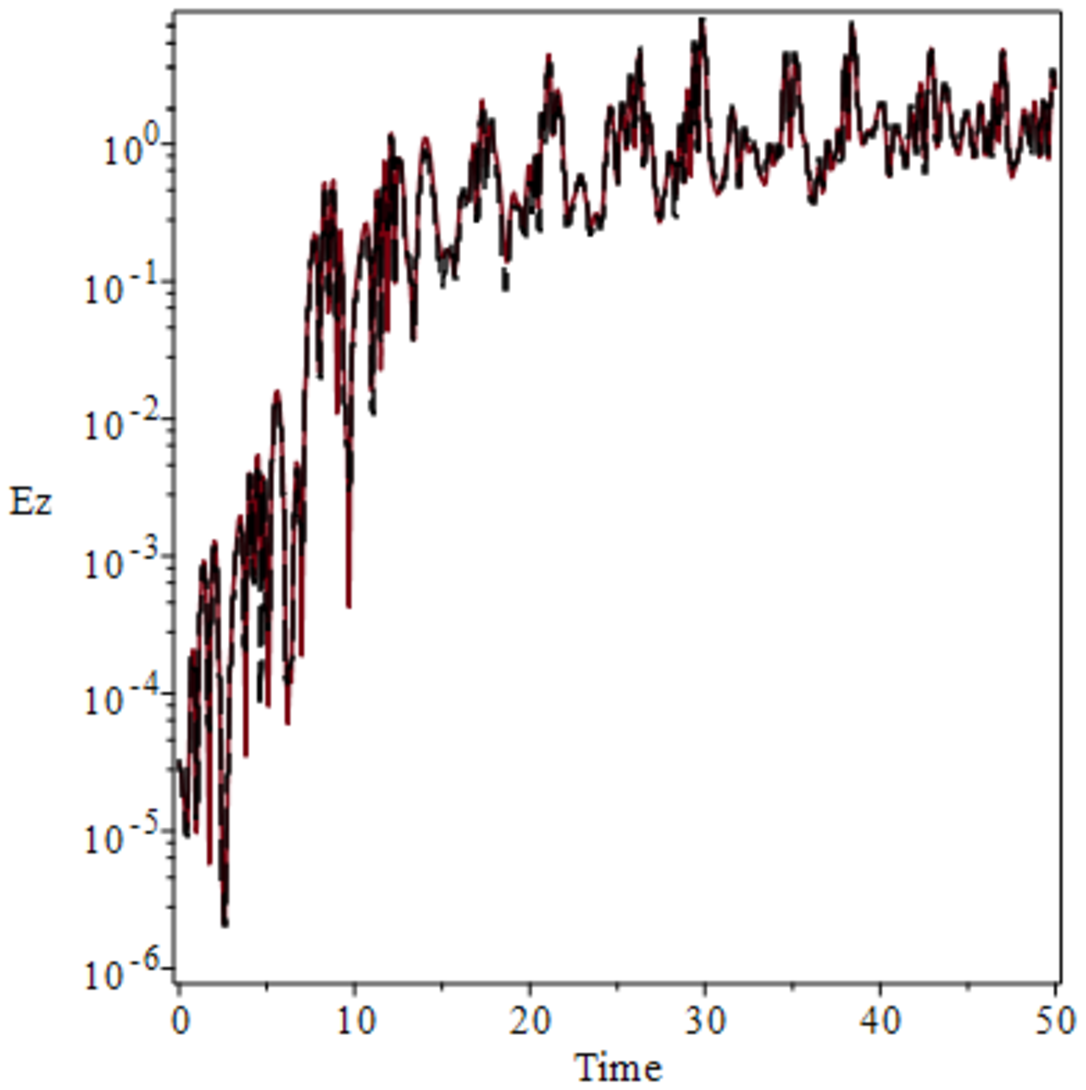}
	\caption{Error growth analysis using 250 perturbed scenarios, with comparable accuracy for the GWRM and RK4. GWRM: \textit{brown line}, $m=8$, $tol=10^{-5}$ and $\epsilon=10^{-5}$. RK4: \textit{black dash line}, steps=15000.}
	\label{fig:1}
\end{figure}

The most evident advantage that GWRM has over RK4 is that the time sub-intervals are typically two orders of magnitude larger. The time-adaptive scheme also allows for longer time intervals in smooth regions and shorter intervals in high gradient regions. This reduces the computation time for the method. The error growth in these studies show that the Lorenz equations are highly sensitive to perturbed initial conditions, and this will have a dominating effect where weather predictions are concerned, as is well established. This shows the need for a numerical model that is efficient at higher accuracies. As can be seen from FIG. \ref{fig:1}, RK4 and GWRM give very similar error growths for the same accuracies. It is also seen that the characteristic time $T$ before which predictions are valid is about $T=10$ time units, whereafter the error growths tend to saturate.

\section{Discussion}
The GWRM is a time-spectral method and differs substantially from traditional time-stepping methods. In this study a first assessment is made of its potential for modelling advanced NWP problems by addressing the Lorenz 1984 chaotic equations. The results are only tentative with respect to NWP, but motivate further studies where space dependence is taken into account. Of particular interest here is that recent development of the GWRM for problems in magnetohydrodynamics has lead to algorithms where high efficiency in handling the spatial dimensions have been obtained, using subdomains. The physical equations of each spatial subdomain may here be solved locally whereas the internal boundary condition equations are solved globally. This enables parallelization of the temporal domain.

An interesting feature of the GWRM is the possibility to average over the fast time scale for problems with several time scales \cite{Scheffel:GWRM1}. A consequence is enhanced efficiency, since only lower order Chebyshev expansions are needed. This approach would be of considerable interest when performing multiple perturbation analysis of a base scenario because fewer scenarios would have to be computed, enhancing efficiency. For the present strongly nonlinear problem, however, high accuracy was found necessary for convergence to the correct solution. The multiple roots of the nonlinear algebraic system of equations (13) are closely distributed. Thus SIR easily departs from the correct solution, corresponding to the initial conditions given, for low temporal resolution. An effort to circumvent this tendency was made by linearising equations (1)-(3) before solution using SIR and gradually adding the nonlinear terms during the iterations. This approach is again not successful because of the tendency of root solvers to lock at the positions of the roots corresponding to the linear equations when the nonlinear terms are included. 

Another question of interest is whether very long time intervals ($>>1$ time unit) could be employed, using the GWRM. The lesson learned from various problems solved so far is that the manageable interval length is indeed problem dependent. For non-smooth problems like the present, the time interval length is limited even if the Chebyshev order is increased because, again, of the spurious solutions found.

Although a causal problem is solved, the GWRM need not necessarily be used as a causal method since both initial and end conditions may be applied. This is an interesting possibility for time-spectral methods which, however, requires further theoretical understanding. Various combination of initial and end conditions have indeed been tried, all within the attractor domains of the solutions, but no improvement of convergence was found.  

Usually overlapping time intervals, employing two point contact, improves convergence. It is notable, and worthy of further study, that the best results (maximum efficiency) were here obtained using one point contact.

Summarizing, the fact that the GWRM competes well with standard finite difference methods in solving the demanding Lorenz 1984 equations and that very efficient GWRM techniques have been developed for pde's where spatial dimensions are included, suggests that usage of time-spectral methods for NWP is well worth further exploration.

\section{Conclusion}
In this work, a preliminary evaluation of a recently developed time-spectral method GWRM is carried out with respect to potential use for numerical weather prediction. The Lorenz (1984) chaotic equations are solved. The efficiency and the behaviour of the error growth with time is compared to traditional explicit and implicit finite difference methods. In particular, the optimal length of the solution time intervals have been determined, the accuracy of the method has been studied in detail and predictability has been investigated. 

It is found that GWRM efficiency is in parity of, or better than, the finite difference methods. Efficiency is further enhanced for cases where several perturbed scenarios need be computed. This is mainly due to that GWRM time intervals are two orders of magnitude larger than those of finite difference methods. Furthermore, the GWRM solutions are analytical Chebyshev series expansions. These findings, and the existence of efficient algorithms for time parallelisation of spatially dependent pde's, are encouraging for future studies of time-spectral methods for NWP. 

\section{Acknowledgement}
\noindent The authors would like to thank Dr {\AA}ke Johansson at SMHI (Swedish Meteorological and Hydrological Institute) for inspiration and for pointing at the relevance of solving the Lorenz equations for a primary evaluation of the GWRM as a method for NWP.

\label{}

%% The Appendices part is started with the command \appendix;
%% appendix sections are then done as normal sections
%% \appendix

%% \section{}
%% \label{}

%% References
%%
%% Following citation commands can be used in the body text:
%% Usage of \cite is as follows:
%%   \cite{key}         ==>>  [#]
%%   \cite[chap. 2]{key} ==>> [#, chap. 2]
%%

%% References with bibTeX database:
\section*{References}
\bibliographystyle{elsarticle-num}
\bibliography{mybib}

%% Authors are advised to submit their bibtex database files. They are
%% requested to list a bibtex style file in the manuscript if they do
%% not want to use elsarticle-num.bst.

%% References without bibTeX database:

% \begin{thebibliography}{00}

%% \bibitem must have the following form:
%%   \bibitem{key}...
%%

% \bibitem{}

% \end{thebibliography}

\end{document}